\documentclass[reprint,prl,aps]{revtex4-2} 

\usepackage{dcolumn}
\usepackage{todonotes} 
\usepackage[ breaklinks=true]{hyperref}
\usepackage{graphicx}
\usepackage{subcaption}
\usepackage{booktabs}
\usepackage{float}
\usepackage[normalem]{ulem}
\usepackage{bm}
\usepackage{amsmath}
\usepackage{amssymb}

\usepackage{comment}
\usepackage{tikz}
\usetikzlibrary{calc} 
\usepackage{upgreek}
\usetikzlibrary{shapes, arrows.meta, 3d, positioning}

\def\O{\mathcal{O}}
\def\u{\mathbf{u}}
\def\x{\mathbf{x}}

\def\q{\mathbf{q}}
\def\n{\mathbf{n}}
\def\r{\mathbf{r}}

\begin{document}

\preprint{APS/insert.number}

\title{Viscous Settling of Bravais Unit-Cells}

\author{Sebastian B\"urger}
\email{sebastian.burger@mis.mpg.de}
 \affiliation{Max-Planck-Institute for Mathematics in the Sciences, 04103 Leipzig, Germany}
 
\author{Harshit Joshi}%
 \email{harshit.joshi@icts.res.in}
\affiliation{%
Univ. Bordeaux, CNRS, LOMA, UMR 5798, 33405 Talence, France 
}%

\author{S Ganga Prasath}
 \email{sgangaprasath@smail.iitm.ac.in}
\affiliation{
 Department of Applied Mechanics, IIT Madras, Chennai 600036, India
}%
\author{Rahul Chajwa}
 \email{rchajwa@stanford.edu}
\affiliation{%
 Department of Bioengineering, Stanford University, Stanford, CA 94305, USA
}%

\author{Rama Govindarajan}
 \email{rama@icts.res.in}
\affiliation{%
  International Centre for Theoretical Sciences, Shivakote Bengaluru 560089, India
}%

\date{\today}

\begin{abstract}

We study  experimentally and theoretically  the Stokesian settling  of a well-known class of porous shapes:  Bravais lattice unit-cells, whose porosity we vary controllably  by changing  their lattice spacing. In our experiments, conducted in a square cuboidal container with its long-axis aligned along gravity, we find that the settling speed $U$ and the solid fraction $\phi_s$ of these lattice units obey a power-law relationship $U \propto \phi_s^{\gamma}$, with an exponent $\gamma = 0.43$ independent of their shape. To  understand the observed scaling exponent, we  analytically and numerically investigate the settling of the simple cubic structure under different approximations. 
We find that the walls of the container, though far from the sinking object, have a defining effect. Our Stokesian  boundary integral simulations show that the Fax\'en boundary correction captures the wall-effects accurately and enables us to discount the wall-effect from the experimental data, yielding a power-law exponent $\gamma = 0.30$ for settling in an unbounded domain. The power-law relating sinking speed and porosity is a step towards predictively understanding the sedimentation fluxes of complex objects in the clouds and the oceans. However, the applicability of this universal scaling to irregular and biologically richer aggregates found in nature remains an open direction.

\end{abstract}


\maketitle
\section{Introduction}
\noindent Sedimentation of particles through viscous fluids is ubiquitous in geophysical  and industrial processes~\cite{GuazzelliMorris2011, Clift2005, zhao2025distribution}, such as the settling of marine snow\cite{Alldredge1988}, microplastics \cite{DiBenedetto2026} and other biogenic particles \cite{Boyd2019} in the oceans, precipitation in reacting flows \cite{Nabika2019}, and the sedimentation of ice crystals in cirrus clouds \cite{Mitchell2008}. The limited predictability of settling speeds of particles in the ocean \cite{LaurenceauCornec2019} and the atmosphere \cite{Mitchell2008} leads to significant uncertainties in climate models. In the limit of small Reynolds number, the motion of isolated rigid particles is governed by the Stokes equation \cite{KimKarrila2005, HappelBrenner1983}, where geometric effects enter through boundary conditions imposed on the particle surface, and the settling velocity is determined by a balance between gravitational forcing and viscous drag. 

\begin{figure*}[t]
\centering
\includegraphics[width=0.9\textwidth]{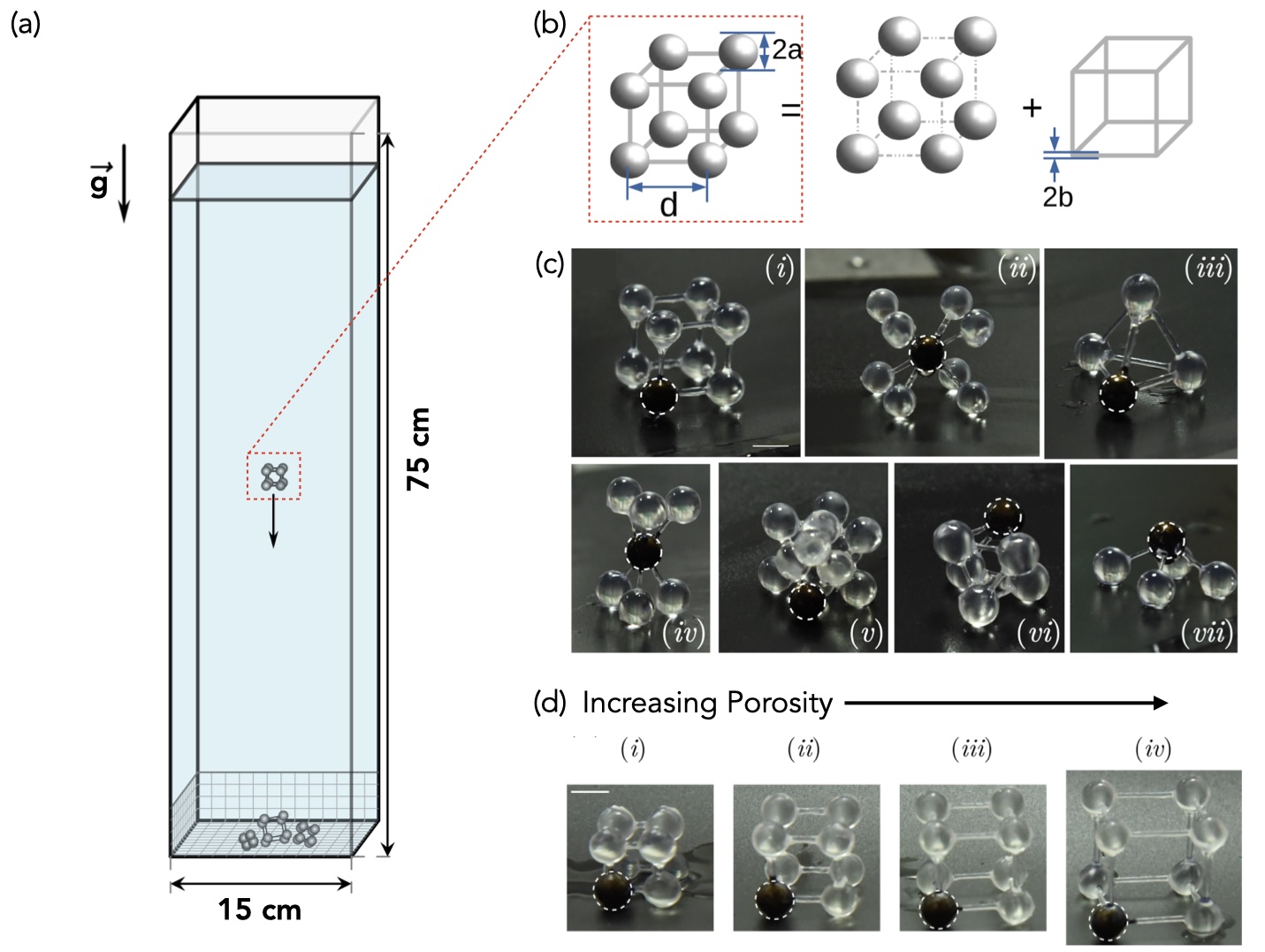}
\caption{$(a, b)$ Schematic of the experimental setup and the settling porous object: an Simple Cubic lattice made of spheres of diameter $2a=6\ \mathrm{mm}$ which are connected by cylindrical rods of diameter $2b=1\ \mathrm{mm}$. The centre-to-centre distance between spheres is $d$. $(c)$ Seven Bravais lattice structures we investigate in our sedimentation experiments: (i) Simple cubic (SC), (ii) body-centred cubic (BCC), (iii) tetrahedron, (iv) hexagonal close-packed (HCP), (v) face centred cubic (FCC), (vi) octahedron and (vii) square-based pyramid structures. Each of them contains one black-painted sphere to track the settling path, speed and orientation. For BCC and FCC, the lattice constant $d$ refers to the unit cell size and not to the nearest centre-to-centre distance of spheres. $(d)$ Sequence of SC-lattices with increasing lattice constant $d$, from left to right in $\mathrm{mm}$: 8, 10, 12, 16. The scale bar is 5  $\mathrm{mm}$.}
\label{fig_struct_pics}
\end{figure*}

In many cases, settling objects possess an internal structure that allows fluid to pass through them. This internal geometry-dependent resistance is 
often described through lumped quantities such as porosity ~\cite{bear2013dynamics}. A common approach is to describe such objects as effective porous media, in which the flow is governed by Darcy's law or its viscous extension, the Brinkman equation \cite{brinkman1947}. In this framework, the fluid motion within the porous aggregate is governed by a linear relationship between the pressure gradient and the filtration velocity, effectively treating the microstructure as a uniform resistive medium. Darcy's law  \cite{brinkman1947} and its later variants 
have been used to estimate drag and settling velocities of permeable particles, beginning with the classical work on sedimentation of polymers \cite{Debye1948}, and developed further for the analyses of 
slow bulk flow through stationary low porosity beds \cite{HappelBrenner1983}. However, these studies do not account for boundary effects from the no-slip conditions and viscous shear at the outer boundary of a non-spherical settling object. Subsequent  models based on the Brinkman equation allow for the matching of interior and exterior flow fields but introduce effective parameters like permeability which must be supplied, in the form of an empirical relationship with porosity \cite{Starov2001}.
An alternative approach treats porous object as a collection of particles interacting hydrodynamically. Through pairwise addition of forces and torques, building on the two-body solution \cite{Stimson1926}, this approach has been used to study many-particle systems, enabling the calculation of close-packed conglomerates \cite{Cichocki1995, Lasso1986}, periodic array of spheres \cite{Sangani1982}, and random or fractal aggregates \cite{Alabrudzinskietal2009, binder2012settling}. In these systems, relationships between settling velocity and porosity or fractal dimension have been proposed for specific models and geometries, but it remains unclear how universal these relationships are across geometry of finite--porosity objects.

In the present work, we study the coupling between flow and geometry of a sinking porous object, using a minimal system which allows for precise control of porosity through a single tunable parameter. We experimentally study low Reynolds number sedimentation and flow of 3D printed unit cells of Bravais lattices \cite{Kittel1954} with equal sized spheres at the lattice nodes. The spheres are connected by slender rods to create rigid ball-and-stick structures. By varying the lattice spacing, the porosity of the structure is tuned systematically while preserving its symmetry. We experimentally find that the settling speed $U$ as a function of the solid volume fraction $\phi_s$  follows a power-law scaling $U \propto \phi_s^{\gamma}$, when $U$ is appropriately rescaled. Remarkably, this power law exponent $\gamma = 0.43$ is measured 
across the different unit-cell structures, suggesting an underlying universality in the settling behavior. 

To understand these observations, we conduct a detailed theoretical analysis of one of these shapes, the simple cubic, 
using a hierarchy of models of increasing fidelity, including analytical approximations based on Stokeslet interactions and slenderbody theory \cite{KimKarrila2005}, as well as numerical simulations using Stokesian Dynamics (SD) \cite{brady1988stokesian} and Boundary Integral Methods (BIM) \cite{Pozrikidis1992}.

These calculations reveal that the presence of confining boundaries has a significant effect on the measured settling velocity, even when the container dimensions are large compared to the particle size.

We show that Fax\'en boundary corrections \cite{HappelBrenner1983} are sufficient for eliminating this effect from the experimental data, allowing us to infer the corresponding unbounded behaviour, which is geophysically relevant. After accounting for wall effects, the power-law exponent reduced to $0.30$, a scaling that is independent of the internal geometry of the object in an unbounded fluid.

\section{Experiments}
\label{sec_exp}
\subsection{Experimental setup}
The settling experiments were conducted in a 75 cm high glass tank with a square cross-section of width 15 cm. The settling particles were fabricated in a Formlabs stereolithography printer, with resin density $\rho_{s} = 1.164$ g/cc. Each object represents a member of a Bravais-lattice unit cell, with equal sized spheres of diameter $2a = 6$ mm placed at the lattice nodes, connected by slender structures (rods) of thickness $2b$ to ensure rigidity of the unit-cell during sedimentation. We consider seven kinds of Bravais structures: simple cubic (SC), body centred cubic (BCC), tetrahedron, hexagonal closed packed (HCP), face centred cubic (FCC), octahedron and square-based pyramid (see Fig.~\ref{fig_struct_pics}(c)). We tune the solid-fraction $\phi_{s}=1-\phi$, with $\phi$ being the porosity, of each Bravais class, by printing the same structure with varying lattice parameter $d$. This presents a controlled class of porous geometries that enables systematic investigation of how microstructure determines settling dynamics. Detailed information about the individual structures is provided in Table \ref{Tab_Strcuture_Data}. Note that $\phi_{s}$ of each object is calculated using its convex hull, as detailed in appendix \ref{sec_steiner_formula}.

\begin{table}[h]
\centering
\begin{tabular}{@{}lccccccc@{}}
\toprule
\textbf{Structure} & \textbf{$N_s$} & \textbf{$N_c$} & \textbf{$N_\mathrm{lattice}$} & \textbf{$d_\mathrm{min}$} & \textbf{$d_\mathrm{max}$}& \textbf{$\phi_\mathrm{s,min}$}  & \textbf{$\phi_\mathrm{s,max}$}\\ \midrule
SC                 & 8 & 12 & 12 & 6 & 20 & 0.06 & 0.61 \\
BCC                & 9 & 8 & 6 & 7 & 18 & 0.08 & 0.53 \\
Tetrahedron        & 4 & 6 & 7 & 6 & 18 & 0.15 & 0.71 \\
HCP                & 7 & 6 & 2 & 6 & 10 & 0.24 & 0.56 \\ 
FCC                & 14 & 24 & 3 & 8.5& 14 & 0.22 & 0.58 \\
Octahedron         & 6 & 12 & 3 & 6 & 14 & 0.17 & 0.69 \\
Sq. Pyramid        & 5 & 4 & 3 & 10& 14 & 0.19 & 0.33 \\
\bottomrule
\end{tabular}
\caption{List of geometric properties of the seven lattice structures depicted in Fig.~\ref{fig_struct_pics}c.  $N_s$ is the number of spheres, $N_c$ the total number of rods connecting the spheres. $N_\mathrm{lattice}$ indicates the number of different lattice constants investigated, with $d_\mathrm{min}$ and $d_\mathrm{max}$ being the minimal and maximal lattice constant, in $\mathrm{mm}$. $\phi_\mathrm{s,min}$ and $\phi_\mathrm{s,max}$ represent the structural solid volume fraction calculated at $d_\mathrm{max}$ and $d_\mathrm{min}$ respectively. For SC, tetrahedron, HCP and octahedron, the lattice constant refers to the nearest neighbour distance. For BCC, FCC and the square pyramid, $d$ signifies the distance of neighbouring spheres that form the large bottom square, see Fig. \ref{fig_struct_pics}c. 'Distance' refers to 'centre-to-centre-distance'.}
\label{Tab_Strcuture_Data}
\end{table}

The glass container was filled with polydimethylsiloxane fluid (silicone oil) of  kinematic viscosity $\nu = 6000$ cSt and density $\rho_{f} = 0.97$ g/cc. Each unit-cell structure was gently submerged and reoriented in the fluid using forceps and needle to avoid any bubble entrainment and prepared to freely sink as close to the centre-line of the container as possible, with one of the unit-cell facing parallel to the front plane at all times. Based on the single sphere diameter $6$ mm, the Reynolds number, Re $=\rho_{f}U a/\mu$  is $\O(10^{-4})$, thus the dynamics takes place in the Stokesian regime. The settling dynamics was captured using a Nikon D700 DLSR camera with image frame-rate $1/3 \ \mathrm{Hz}$. One sphere in each object was coloured black (see Fig.~\ref{fig_struct_pics} (c)\&(d)) to help with tracking. The images were converted to grayscale and the object was tracked in a MATLAB image analysis pipeline, yielding position as a function of time. The sedimentation velocity was measured when the particle was approximately half way in its descent, away from the top and bottom surfaces, for a short time-scale in which both orientation and velocity were measured to be nearly constant. However, we do measure residual rotation of the object during the time of the full trajectory, which can be attributed to the boundaries.

\begin{figure*}[t]
    \centering
        \includegraphics[width=\linewidth]{figs/Figure2.jpeg}
    \caption{(a) Schematic of the setup of the PIV-experiment with laser-sheet bisecting the object, revealing the flows in the mid-plane between the four spheres in the front and their counterparts in the background. The extraction line indicates where the velocity profile in the PIV evaluation is measured. (b) Front-view of the PIV snapshot for SC structure with $d=10 \ \mathrm{mm}$, with suspended tracer beads in the background. Note that the PIV plane is different from the plane of the four visible cylinders, as Fig (a) indicates. (c)  Streamlines from PIV images of the mid-plane showing circulation zones close to the boundary. The color depicts the magnitude of vertical component of the fluid velocity.}
    \label{piv_setup}
\end{figure*}

Beyond individual particle trajectories, we measured the two-dimensional slice of the fluid flow to gain insights into the fluid-structure interaction within the sinking structure. Tracer glass beads were suspended in the fluid and homogenized using a coiled wire attached to a hand-held kitchen mixer. Note that homogenizing tracer particles/powder in a large volume of a highly viscous fluid is challenging. Even after the tracer particles were suspended as uniformly as possible using the mixer, some inhomogeneities arise from entrainment when we retrieve the objects out of the fluid, as shown in Fig. \ref{piv_setup}(b), however their effect on measured velocity field is suppressed by the coarse-graining involved in PIV analysis. We performed Particle Image Velocimetry (PIV) by shining the tracers using a laser beam (\textsc{Wicked lasers} 300 mW, $\lambda = 532\ \mathrm{nm}$). The quasi-two-dimensional laser sheet was formed by the beam passing through a cylindrical lens, with an adjustable position such that the laser sheet bisected the sinking object, illuminating the tracer flows in its mid-plane, as visible in Fig.~\ref{piv_setup}. The recorded movies of the settling structures with tracer particles were analysed using \textsc{Pivlab} to produce the flow field [see Fig. \ref{piv_setup}(c)].

\begin{figure}[t]
    \centering
        \includegraphics[width=\linewidth]{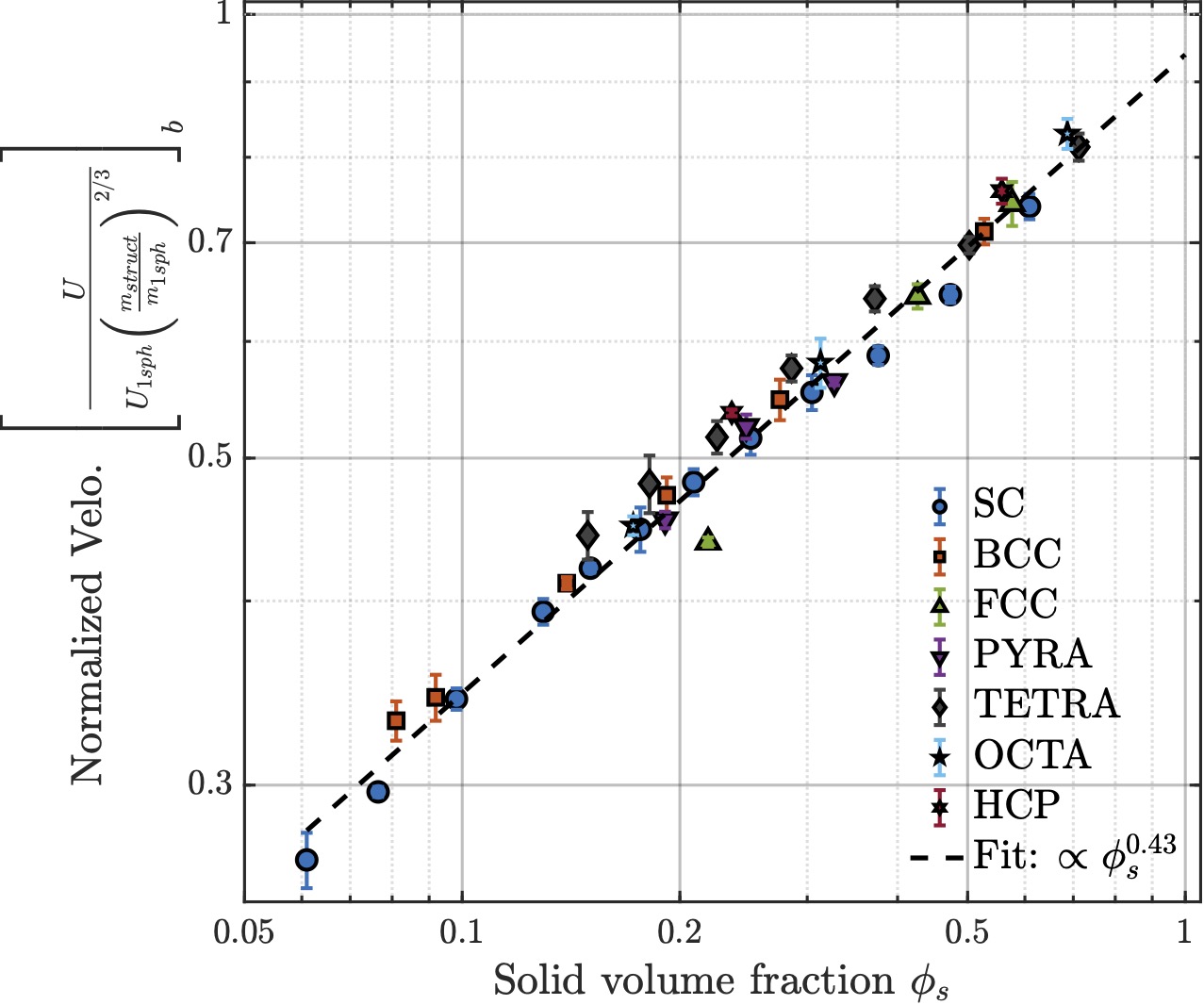}
    \caption{Normalized terminal velocity obtained in the various experiments as a function of the structures solid volume fraction $\phi_s$, where the porosity is $(1-\phi_s)$. A universal scaling across the different structures is visible, with a power-law fit $U \propto \phi_s^{\gamma} \ $  yielding an exponent $\gamma=0.433 \pm 0.004$.  As described using eqn.~\eqref{scaling_exp_2_3}, the velocity is normalized against the velocity of a sphere of the same mass, $U_\mathrm{1sph}(m_{\mathrm{struct}}/m_{\mathrm{1sph}})^{2/3}$. Subscript 'b' indicates  the settling speeds as  measured in the container, i.e. a bounded domain.}
    \label{fig_raw_power_law}
\end{figure}

\subsection{Scaling analysis}
In the well-known case of a single sphere steadily settling in an unbounded domain in Stokes flow, drag as given by Stokes' formula is balanced by gravity:
\begin{align}
      6 \pi \mu a  U&{=}    \frac{4}{3} \pi a^3 \Delta \rho \  g,
    \label{sixpi}
    \\
   U &= \frac{2}{9}  \frac{\Delta \rho \ g}{\mu } a^2 \sim m^{2/3},
    \label{scaling_exp_2_3}
\end{align}
with  $g$ being gravitational acceleration, $\Delta \rho = \rho_{p} - \rho_{f}$ is the excess density of the solid sphere over that of the fluid, and $m$ the mass of the sphere. Equation~\eqref{scaling_exp_2_3} determines the influence of mass onto the settling speed of a single sphere. For the bravias unit-cell structures, the Stokes' settling speed depends on both the geometry and the mass, wherein the shape effects are captured by a lumped parameter -- porosity. In order to isolate the effect of porosity onto the structure's settling speed $U$ from the mere effect of its mass $m_\mathrm{struct}$ , we normalize  it by $m_\mathrm{struct}^{2/3}$ (see Fig.~\ref{fig_raw_power_law} and \ref{fig_porosity_vs_velocity}), i.e. $U$ is rescaled with $(m_\mathrm{struct}/m_\mathrm{1sph})^{2/3} U_\mathrm{1sph}$. Note that throughout this paper the subscript $\mathrm{1sph}$ will refer to a single sphere, of diameter $2a=6 \ \mathrm{mm}$, either in isolation or as part of a structure. Accordingly, $U_\mathrm{1sph}$ is the settling speed of such a sphere.  For example, in the case of 8 spheres of radius $a$ without rods, arranged in an sc lattice, the normalization factor is $(m_\mathrm{struct}/m_\mathrm{1sph})^{2/3}U_\mathrm{1sph}= 4 \ U_\mathrm{1sph}$,
being the velocity of a single sphere with the combined mass of these 8 spheres. In summary, this normalization enables us to compare the effect of porosity on the settling speeds of structures with different mass, as can be seen in Fig. \ref{fig_raw_power_law}. 
\\

In the following sections 3 \& 4, we lay the analytical and numerical framework, respectively, and then compare the experiments and theory in section 5.

\section{Theoretical framework} 
\label{section_math}

In our experiments, we study seven different Bravias lattice structures and evaluate their settling speed as a function of the porosity. However, in the theoretical analysis, we confine ourselves to the sedimentation dynamics of the simple cubic (SC) lattice structure alone, as it is amenable to a variety of approximations. As we shall see, the different approximations provide different estimates of the settling speed where the most detailed model which accounts for the wall-effects matches our experimental observations closely. The different approximations we use to capture the SC lattice sedimentation, with increasing level of details, are: (i) assume the structure as a collection of 8 Stokeslets, where each sphere interacts with the other spheres while the connecting rods are ignored, and it is assumed that $a/d \ll 1$; (ii) Stokeslets along with slender-body model for the drag force introduced by the rods, (iii) Stokesian Dynamics (SD) which accounts for near-field lubrication effects and the finite size of the spheres, again for 8 spheres; (iv) SD with the addition of the slender rods made up of an array of small spheres of radius $b$ (see Fig.~\ref{fig_sd_struct}); (v) Boundary Integral Method (BIM) for the set of 8 spheres of radius $a$, settling in a fixed configuration with wall effects.

\subsection{Fax\'en boundary correction}
As we shall see in the results section, we find a considerable mismatch between the experiments, the theoretical and the numerical predictions from unbounded-domain calculations. The missing aspect is the influence of the container walls, i.e., the finite size of the experimental setup, which forces the long-range $1/r$-decaying Stokeslets to zero at the boundary, thus altering the settling speed. 
 The reduced mobility of objects confined within a container can be interpreted in light of the inclusion monotonicity principle, which follows from the minimum-dissipation theorem. Specifically, the principle states that the drag on a particle translating within a rigid stationary container $C_1$ is greater than or equal to the drag on the same particle translating within a container $C_2$, provided that $C_2$ completely encloses $C_1$ \cite{KimKarrila2005, hill1956extremum}. Consequently, accurate prediction of the terminal velocity of structures that settle in a container requires proper accounting of the effects of confinement.

Fax\'en's boundary correction term (see Chapter 7-3 and specifically eqn. (7-3.113) in \cite{HappelBrenner1983}), prescribed for a cylindrical domain of radius $R_{\mathrm{cyl}}$ can be used to calculate the reduced settling speed $U_b$ of a sphere of radius $a$ to be
\begin{align}
U_b & = f(\lambda) \ U_u, \ {\rm where} \\
    f(\lambda) & = 1 - 2.10443\lambda + 2.08877\lambda^3 - 0.94813\lambda^5 \nonumber\\&-1.372 \lambda^6 + 3.87 \lambda^8 -4.19 \lambda^{10},
    \label{Faxen_formula}
\end{align}
with $\lambda = a/R_{\mathrm{cyl}}$. Here, $U_u$ is the speed of the object in an unbounded domain (subscripts $b$ and $u$ will be used for bounded and unbounded hereafter). In order to adapt this to our experiments, two choices have to be made: first, as the structures are not spheres, one needs to define an effective radius $a_\mathrm{eff}$ of the structure in order to be able to apply eqn.~\eqref{Faxen_formula}. To this end, we define $a_\mathrm{eff}$ in analogy with eqn.~\eqref{sixpi} as

\begin{equation}
    F_{G,\mathrm{struct}} = 6\pi\mu a_{\mathrm{eff}} U.
    \label{struct_Stokes_law}
\end{equation}
where the subscript `struct' stands for the unit-cell while the object's settling speed is $U$. Stokes' law \eqref{struct_Stokes_law}  holds for unbounded domains, thus 
\begin{equation}
    a_{\mathrm{eff}} = a \frac{m_\mathrm{struct}}{m_\mathrm{1sph}} \left[\frac{U_\mathrm{1sph}}{U}\right]_u. 
    \label{a_eff}
\end{equation} 

Second, the cross-section of our container is not cylindrical but square, however we choose an $R_{\mathrm{cyl}}:=W/2$, with $W$ being the width of the glass container. Importantly,  Fax\'en's correction  also needs to be applied to our estimates for a single sphere, hence with $\lambda_\mathrm{struct}=a_\mathrm{eff}/R_\mathrm{cyl}$
 \begin{equation}
    \left[\frac{U}{U_\mathrm{1sph}} \right]_b = \frac{f(\lambda_\mathrm{struct})}{f(\lambda_\mathrm{1sph})}  \left[\frac{U}{U_\mathrm{1sph}} \right]_u. 
    \label{velo_inftyy_bounded_conversion}
\end{equation}

\begin{figure*}[t]
    \centering
        \includegraphics[width=0.7\linewidth]{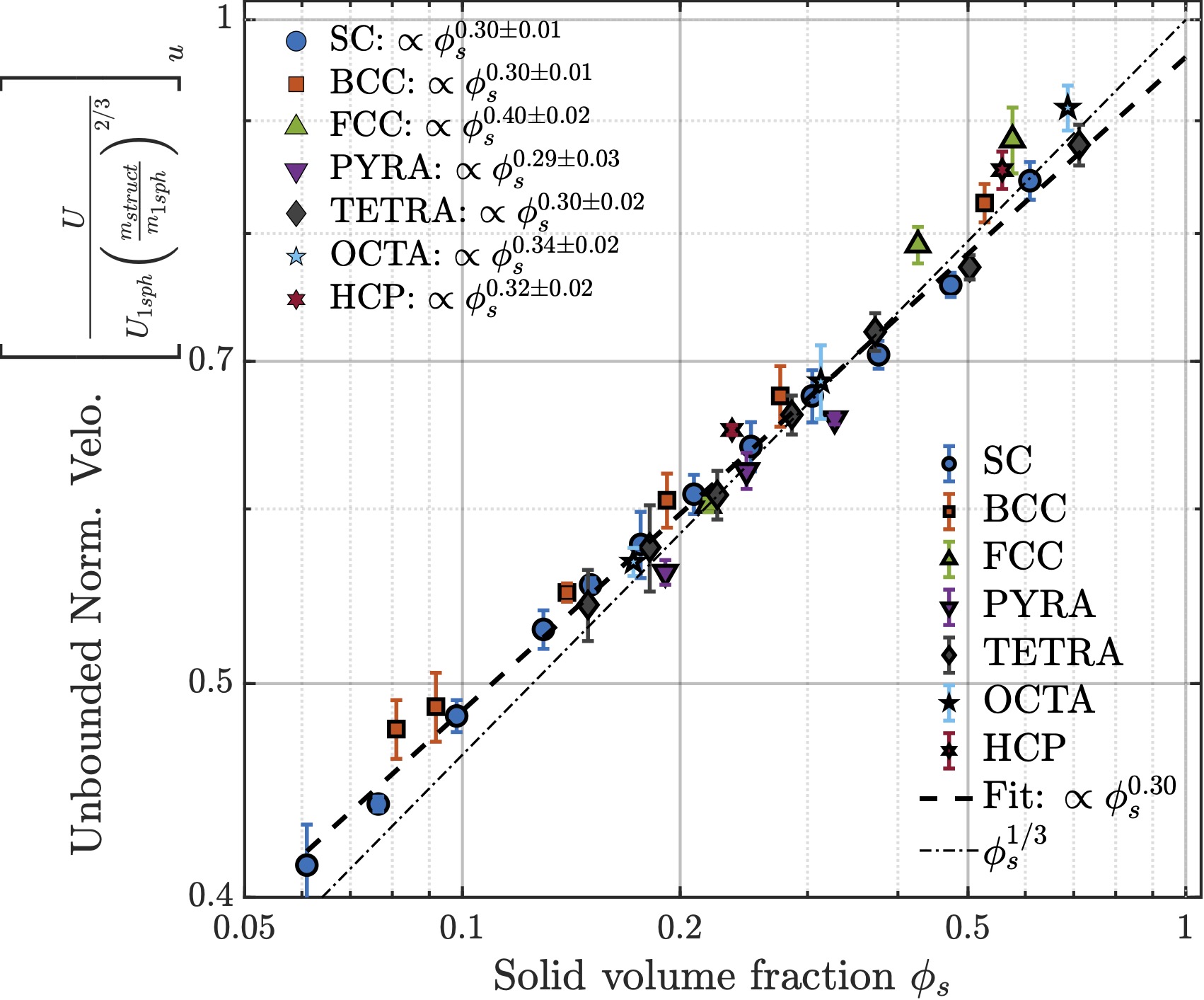}
    \caption{Analogous to Fig.~\ref{fig_raw_power_law}, but with wall effects removed for each data point in the bounded domain by using Fax\'en's boundary correction (subscript 'u' indicating an unbounded domain). The results are now representative of settling in an unbounded domain (such as in the ocean or atmosphere), for which the power law exponent is $\gamma =0.296 \pm 0.004 $. Note that with the exception of the fcc structure, the individual structure type's exponents are in good agreement.}
    \label{fig_porosity_vs_velocity}
\end{figure*}

Thus, eqn. \eqref{a_eff} and \eqref{velo_inftyy_bounded_conversion} together give the conversion of $U_u$ to $U_b$. Note, however, that in the experiment,  bounded quantities are measured, from which we aim to estimate the unbounded ones. Applying  the correction requires solving the implicit eq.~\eqref{velo_inftyy_bounded_conversion} i.e. the calculation of $(U/U_\mathrm{1sph} )_u $ requires $\lambda_\mathrm{struct}= 2a_\mathrm{eff}/W$, which itself  requires $(U/U_\mathrm{1sph} )_u $ via \ref{a_eff}. This is therefore solved numerically, and the results of this bounded-to-unbounded-domain conversion are displayed in Fig. \ref{fig_porosity_vs_velocity}. \\

\subsection{Stokeslet approximation with and without slender rods}
\label{sec_Stokeslet_analytical}

Consider a collection of $n$ equal force monopoles of strengths $\bf{F}$, located at positions ${\bf r_{1}, r_{2},..,r_{n}}$, and interacting with each other hydrodynamically. In the limit of small $a/d$, where $a$ is the size a single particle and $d$ is the inter-particle separation, the dynamics of $i^{\mathrm{th}}$ particle, to leading orders in $a/d$ (see ch.6 in \cite{HappelBrenner1983}), can be written as
\begin{equation}
     \dot{\bf{r}}_{i} = \frac{\bf F}{6\pi\mu a} + \sum_{j \neq i} \mathbf{F} \cdot \mathbb{G}(\mathbf{r}_{i} - \mathbf{r}_{j}),
\end{equation}
where $\mathbb{G}$ is the Stokesian hydrodynamic kernel \cite{KimKarrila2005}. The hydrodynamic interactions yields a reduced drag \cite{Crowley1971} that makes the collection of particles fall faster than the individual.

By the same token, a simple cubic unit without rods falls faster than each of the spheres would on their own, with a speed
\begin{equation}
 \mathbf{U}_{sc} = \left( 3  + \frac{3}{\sqrt{2}} + \frac{1}{\sqrt{3}}\right) \frac{a \,  \mathbf{U}_\mathrm{1sph}}{d}.
\label{formula_a_i}
\end{equation} 

Note that the lateral drifts from hydrodynamic interactions are cancelled by the symmetry of the unit cell and its orientation with respect to gravity, and the rigidity constrain. We consider the effects of the rods connecting the spheres by adding the drag contribution using the slender-body theory, assuming that the rods do not interact hydrodynamically with each other. In the SC lattice there are eight horizontal and four vertical rods, and total vertical drag force on the rods is
\begin{equation}
 \mathbf{F}^{|-}_{sc} = 8  \mathbf{F}_{\perp} \, + \, 4  \mathbf{F}_{||} \, = \, \frac{40 \pi \mu (d-2a)}{\ln[\frac{d-2a}{b}]}  \mathbf{U}_{|-}.
\label{eqn_rod_drag}
\end{equation} 
Here $\mathbf{F}_{\perp}$ and $\mathbf{F}_{||}$ are the drag forces perpendicular to and parallel to the symmetry axis of a rod. Balancing the drag force $F^{|-}$ with the net buoyant weight of the rods gives $\mathbf{U}_{|-}$. As the lattice spacing $d$ increases, so does the contribution of the rods, and their effect on the dynamics become important in the settling velocity.

Distributing the total drag force from \eqref{eqn_rod_drag} among the eight spheres in the SC unit cell yields the resistivities 
 \begin{equation}
     \beta_\mathrm{1sph}=6\pi \mu a  \quad \mathrm{and} \quad \beta_\mathrm{rod}=\frac{5 \pi \mu (d-2a)}{\ln[\frac{d-2a}{b}]},
 \end{equation}
hence the resistance-weighted average amounts to 
\begin{equation}
    \bar{\beta}_\mathrm{1sph}=\frac{\beta_\mathrm{1sph}}{\beta_\mathrm{1sph} + \beta_\mathrm{rod}} \ = \ \left[1 + \,\frac{5}{6} \,\frac{d/a- 2}{\ln \left[\frac{d-2a}{b}\right]}\right]^{-1},
\end{equation}  
The settling velocity of the SC unit-cell accounting for the rods thus becomes
\begin{align}
\frac{U_{sc}}{U_\mathrm{1sph}}  &=  \bar{\beta}_\mathrm{1sph} + \left[1 -  \bar{\beta}_\mathrm{1sph}\right]\, \frac{27}{20} \left(\frac{b}{a}\right)^2 \ln \left[\frac{d-2a}{b}\right] \nonumber\\ &+  \left(3  + \frac{3}{\sqrt{2}} + \frac{1}{\sqrt{3}}\right) \frac{a}{d}.
\label{formula_a_ii}
\end{align}

Here we have redistributed the drag due to the rods on to the spheres and this is possible for the SC structure given its symmetry. This special symmetry is not available for other structures, where we must treat different spheres individually for this purpose.

The Stokeslet analysis is valid in the far-field limit i.e. when $a/d \ll 1$, but for some of the structures we study $a/d$ is $\mathcal{O}(1)$. To capture the full experimental detail with its near-field interactions we conduct numerical simulations using two standard techniques in low Reynolds number hydrodynamics -- Stokesian Dynamics (SD) \cite{brady1988stokesian, Townsend_Stokesian_Dynamics_in_2024} and Boundary Integral Method (BIM) \cite{Pozrikidis1992}.
In addition to making our theoretical analysis more precise, our simulations provide a rigorous justification for using the Fax\'en boundary correction, which will be discussed in section \ref{results}. In the following section we present our numerical scheme in detail.

\section{Numerical simulations}
\label{sec_numerics}
In addition to the analytical estimates above, our experiments on the simple cubic structure are accompanied by Stokesian Dynamics (SD) and Boundary Integral Method (BIM) simulations. Both methods are used within the creeping flow regime ($Re \ll 1$) and the mathematical details will be described in Sections~\ref{numerics_SD} and \ref{numerics_BIM}. In short, BIM relies on expressing the velocity of the structure as an integral over the structure's and container's boundary involving Stokeslet $\bm{\mathcal{G}}$, Rotlet $\bm{\mathcal{R}}$ and Stresslet $\bm{\mathcal{T}}$ which can be solved by defining a triangular mesh on the boundary \cite{KimKarrila2005, Pozrikidis1992, Pozrikidis2002, Prosperetti2009}.
Stokesian Dynamics \cite{brady1988stokesian}, however, does not require a boundary mesh, but models the investigated structure as consisting of a finite number of spheres, represented as point singularities located at the sphere centre. Its interactions are calculated through  a multipole expansion of the force distribution, using $\bm{\mathcal{G}}$, $\bm{\mathcal{T}}$, $\bm{\mathcal{R}},$ etc. Since the multipole expansion captures the physics accurately only in the far-field, we use a two-body lubrication theory to model the fluid flow close to the spherical particles. 

In the experiments, the spheres are connected by rigid rods that restricts relative motion between them, so that the structure settles as a rigid body with a common translational velocity. In the simulations, however, the structures are represented as collections of disjointed spheres, and the rigidity constraint must therefore be imposed explicitly to eliminate relative motion between the spheres. This can be accomplished using the resistance matrix of the assembly, obtained from SD or BIM. Because the constraint forces acting on individual spheres are internal, their sum vanishes. Consequently, the net external force on the structure is simply the sum of buoyancy-corrected weights of the spheres, denoted by $\bm F_{\mathrm{struct}}$. It follows that
\begin{equation}
    \label{eq:commonVel}
    \bm F_{\mathrm{struct}} = \bm \sum_{\alpha,\beta=1}^N \bm{\mathcal{R}}^{tt}_{\alpha\beta}\cdot \bm U,
\end{equation}
where $\bm U$ is the common settling velocity of the structure, $N$ denotes the number of spheres used to model the structure, and $\bm{\mathcal{R}}^{tt}_{\alpha\beta}$ is the translational part of the resistance tensor relating spheres $\alpha$ and $\beta$. Equation \eqref{eq:commonVel} is used to obtain $\bm U$, given $\bm F_{\text{struct}}$. 
In the BIM simulations, the rigid rods connecting the spheres are not included, since the objective is to assess container effects.

\subsection{Stokesian Dynamics}
\label{numerics_SD}

For a structure consisting of $N$ spheres settling in the Stokesian regime, the SD simulation contains a multipole expansion that appropriately captures the far-field interaction of the particles through a grand mobility matrix $\mathcal{M}^\infty$ that determines the velocities $\mathbf{U}^\infty$ arising from the prescribed forces $\mathbf{F}^\infty$ via $\mathbf{U}^\infty =  \mathcal{M}^\infty \cdot \mathbf{F}^\infty$. In our study, we employ the numerical solver in~\cite{Townsend_Stokesian_Dynamics_in_2024} to perform the SD simulations, as it  provided accurate results for the analytically known Stimson-Jeffery solution  (ref.~\cite{Stimson1926}) which is highly relevant for our simulations. SD simulations are performed for SC lattices for different values of the lattice constant $d$ to compare against two experimentally measured quantities: the terminal velocity of the SC structure, and the internal flow profile. The multipole expansion considered in our simulations is truncated at the first moment, i.e. including the stresslet and rotlet. Hence, each sphere is described using three quantities: force, torque and stresslet, determining the sphere's translational and rotational velocity, $\mathbf{U}$ and $\bm{\Omega}$, and its rate of strain $\mathbf{E}$, leading to \cite[eqn. 2.6]{townsend2017mechanics} 
\begin{equation}
    \begin{pmatrix} 
        \mathbf{U}^\infty \\ 
        \bm{\Omega}^\infty \\ 
        -\mathbf{E}^\infty 
    \end{pmatrix} 
    = 
    \begin{pmatrix} 
        \mathcal{M}^{tt} & \mathcal{M}^{tr} & \mathcal{M}^{ts} \\ 
        \mathcal{M}^{rt} & \mathcal{M}^{rr} & \mathcal{M}^{rs} \\ 
        \mathcal{M}^{st} & \mathcal{M}^{sr} & \mathcal{M}^{ss} 
    \end{pmatrix} 
    \begin{pmatrix} 
        \mathbf{F}^\infty \\ 
        \mathbf{T}^\infty \\ 
        \mathbf{S}^\infty 
    \end{pmatrix}.
\end{equation}
Additionally, the mobility matrix includes a second-order operator $\Delta \bm{\mathcal{G}}$ to capture the finite size of the particles. To compute the near-field interaction of the spheres, which becomes especially relevant when we have the slender rods modelled and spheres lying very close to each other, lubrication forces $\mathcal{R}_{\mathrm{lub}}$ for the squeezing and shearing of two spheres are included. Added to  the far-field resistivity $\mathcal{R}=(\mathcal{M}^{\infty})^{-1}  $ and corrected by the far-field contribution $\mathcal{R}_{\mathrm{lub}}^{\infty}$, the total resistivity is
\begin{equation}
    \mathcal{R} = \left(\mathcal{M}^{\infty}\right)^{-1} + \mathcal{R}_{\mathrm{lub}} - \mathcal{R}_{\mathrm{lub}}^{\infty}.
\end{equation}
The resistivity formulation is, in particular, useful for prescribing a constant velocity for a rigid structure via imposing constraint forces. \\

\begin{figure}[t]
    \centering
        \includegraphics[width=0.5\linewidth]{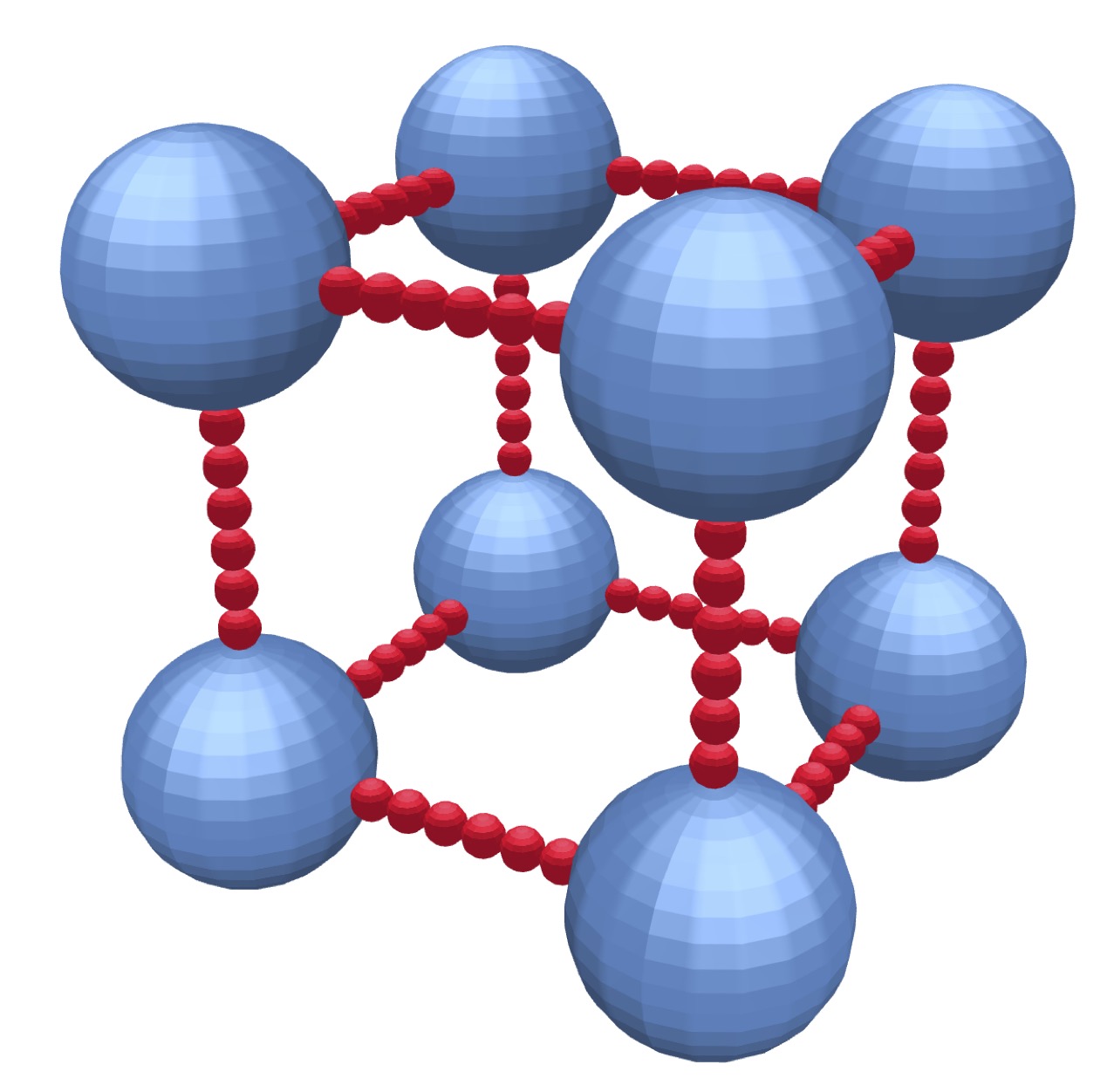}
    \caption{Schematic of the setup used for the Stokesian Dynamics (SD) calculations. The larger blue spheres are of radius $3 \ \mathrm{mm}$ while the string of small red spheres (of radius $0.5 \ \mathrm{mm}$) represent the slender rods. In our analysis, we consider two scenarios: retain the slender rods made of small spheres or omit the rods. The blue spheres are fixed at their relative locations in both cases.}
    \label{fig_sd_struct}
\end{figure}

As the method is limited to spherical particles, we approximate the SC structure as shown in Fig.~\ref{fig_sd_struct} with small red spheres of radius $b=0.5 \ \mathrm{mm}$ and larger blue spheres of  $a=3 \ \mathrm{mm}$. The small spheres represent the slender rods, and are retained in some simulations and omitted in others. When they are retained, we refer to the structure as ``full SC cluster". This approximation of the actual sphere-rod structure then requires two correction factors: (1) Discrepancy in volume between the chain of red spheres ($V_{\mathrm{ch}}$) and an actual cylinder of the same radius ($V_\mathrm{cyl}$): $V_\mathrm{cyl}/V_{\mathrm{ch}}=1.5$; (2) Drag difference between the drag due to a single cylinder (cf.  \eqref{eqn_rod_drag}) and the SD calculated value for an isolated chain of $n(d)$ spheres representing the rod.\\

To obtain the object's velocity $\mathbf{U}$, constraint forces are necessary to enforce rigid-body-motion. The linearity of the Stokes equation allows for the following approach: a trial structure velocity $\mathbf{U}_{\mathrm{guess}}$ is prescribed and subsequently constraint and drag forces are calculated, which needs to balance the known force due to gravity. Rescaling the velocity to satisfy this balance  yields the prediction for $\mathbf{U}$. Further, we find the velocity in the background flow induced by the settling structure by following an extremely small neutrally buoyant particle which behaves as a passive tracer (we prescribe its radius of tracer as $a_T=0.05 \ \mathrm{mm}$). Therefore a mixed boundary value problem arises: the prescribed forces lead to Neumann boundary conditions for the velocity and the constraint forces ensuring uniform settling yield Dirichlet boundary conditions. This results in a system of equations for the constrained structure (S) and the unconstrained tracer (T), using the resistance matrix:
 \begin{equation}
      \begin{pmatrix} \mathbf{F}_S \\ \mathbf{F}_T \end{pmatrix} = \begin{pmatrix} \mathcal{R}_{SS} & \mathcal{R}_{ST} \\ \mathcal{R}_{TS} & \mathcal{R}_{TT} \end{pmatrix} \begin{pmatrix} \mathbf{U}_S \\ \mathbf{U}_T \end{pmatrix}.
      \end{equation}
Since the tracer does not exert any drag, $\mathbf{F}_T=0$, this leads to $\mathbf{U}_T=-\mathcal{R}_{TT}^{-1}\mathcal{R}_{TS}\mathbf{U}_S$  and hence 
\begin{equation}
    \mathbf{F}_S=(\mathcal{R}_{SS}-\mathcal{R}_{ST}\mathcal{R}_{TT}^{-1}\mathcal{R}_{TS})\mathbf{U}_S.
\end{equation}
As $\sum_i \mathbf{F}_\mathrm{S,i} = \mathbf{F}_{G,\mathrm{struct}}$ needs to hold, the same method using rescaling of $\mathbf{U}_\mathrm{guess}$ to obtain $\mathbf{U}$ as used before applies, with a new effective resistance $\mathcal{R}_\mathrm{eff}=(\mathcal{R}_{SS}-\mathcal{R}_{ST}\mathcal{R}_{TT}^{-1}\mathcal{R}_{TS})$.

\subsection{Boundary Integral Formulation }
\label{numerics_BIM}
The boundary integral formulation gives the following representation of the velocity field $\u(\x)$ in the domain $\mathcal{D}$, interior to the container $C$, and  exterior to each of the $N$ rigid bodies $S_\alpha$, indexed by $\alpha=\{ 1,2,..., N\}$ \cite{KimKarrila2005, Pozrikidis1992, Prosperetti2009}:
\begin{multline}
    \u(\x_0) = -\sum_{\alpha=1}^N \frac{1}{4\pi}\oint_{S_\alpha} \q_{\alpha} (\x) \cdot \bm{\mathcal{T}}(\x, \x_0)\cdot \hat{\n}_\alpha dS(\x) \\ +  \frac{1}{4\pi}\oint_{C} \q_{c} (\x) \cdot \bm{\mathcal T}(\x, \x_0)\cdot \hat{\n}_c dS(\x) 
    \\ + \bm{\mathcal V}^s(\x_0), \quad \x_0 \in \mathcal{D},
    \label{velo_BIM}
 \end{multline}
where 
\begin{subequations}
\begin{gather}
\bm{\mathcal{T}} (\x,\x_0) \equiv -\frac{6\r \r \r}{|\r|^5},\quad \r \equiv \x-\x_0, \\
\bm{\mathcal V}^s(\x_0) \equiv \sum_{\alpha=1}^N [-\mathbf{F}^h_\alpha\cdot \bm{\mathcal G}(\x_0,\x_\alpha) +\mathbf{T}^h_\alpha\cdot \bm{\mathcal R}(\x_0,\x_\alpha)], \\
\bm{\mathcal G}(\x_0,\x) = \frac{\bm \delta}{|\r|} + \frac{\r \r}{|\r|^3}, \quad \bm{\mathcal R}(\x,\x_0) \equiv \frac{1}{2} \bm\nabla \times \bm{\mathcal G}(\x_0,\x).
\end{gather}
\end{subequations}
Here $\q_\alpha$ is the double-layer density on the surface of body $\alpha$, $\q_c$ is the double-layer density on the surface of the container, $\bm{\mathcal G}, \bm{\mathcal R}, \bm{\mathcal T}$ are the Stokeslet, Rotlet and Stresslet for a sphere. The supplementary flow needed to complete the double-layer representation is denoted by $\bm{\mathcal V}^s$. 
The hydrodynamic force and torque on body $\alpha$ are denoted by $\mathbf F^h_{\alpha}$ and $\mathbf T^h_\alpha$, respectively. 
Note that $\hat{\n}_\alpha$ denotes the outer unit normal to the body $\alpha$, directed towards the fluid. Similarly, $\hat{\n}_C$ denotes the inner normal to the container, directed towards the fluid and $\x_\alpha$ is the geometric centre of the body $\alpha$. Using the no-slip boundary conditions in equation \eqref{velo_BIM}, one arrives at the following second-kind integral equations for the double-layer densities \cite{KimKarrila2005, Pozrikidis1992, Prosperetti2009}:
\begin{subequations}
\begin{gather}
    \label{eq:qDLeqn}
    (\mathbb{I} + \bm{\mathcal L}^d + \bm{\mathcal P} ^{\text{RB}} + \bm{\mathcal P}^n)(\q \oplus \q_C)[\x_s] =  \nonumber\\ \bigoplus_{\alpha=1}^N\bm{\mathcal V}^s[\x_s \in S_\alpha] \oplus (-\bm{\mathcal V}^s[\x_s \in C]), \\ \q \equiv \bigoplus_{\alpha=1}^N \q_\alpha ,\quad \x_s \in \bigcup_\alpha S_\alpha \cup C.
\end{gather}
\end{subequations}

\begin{figure*}[t]
    \centering
        \includegraphics[width= 0.9\linewidth]{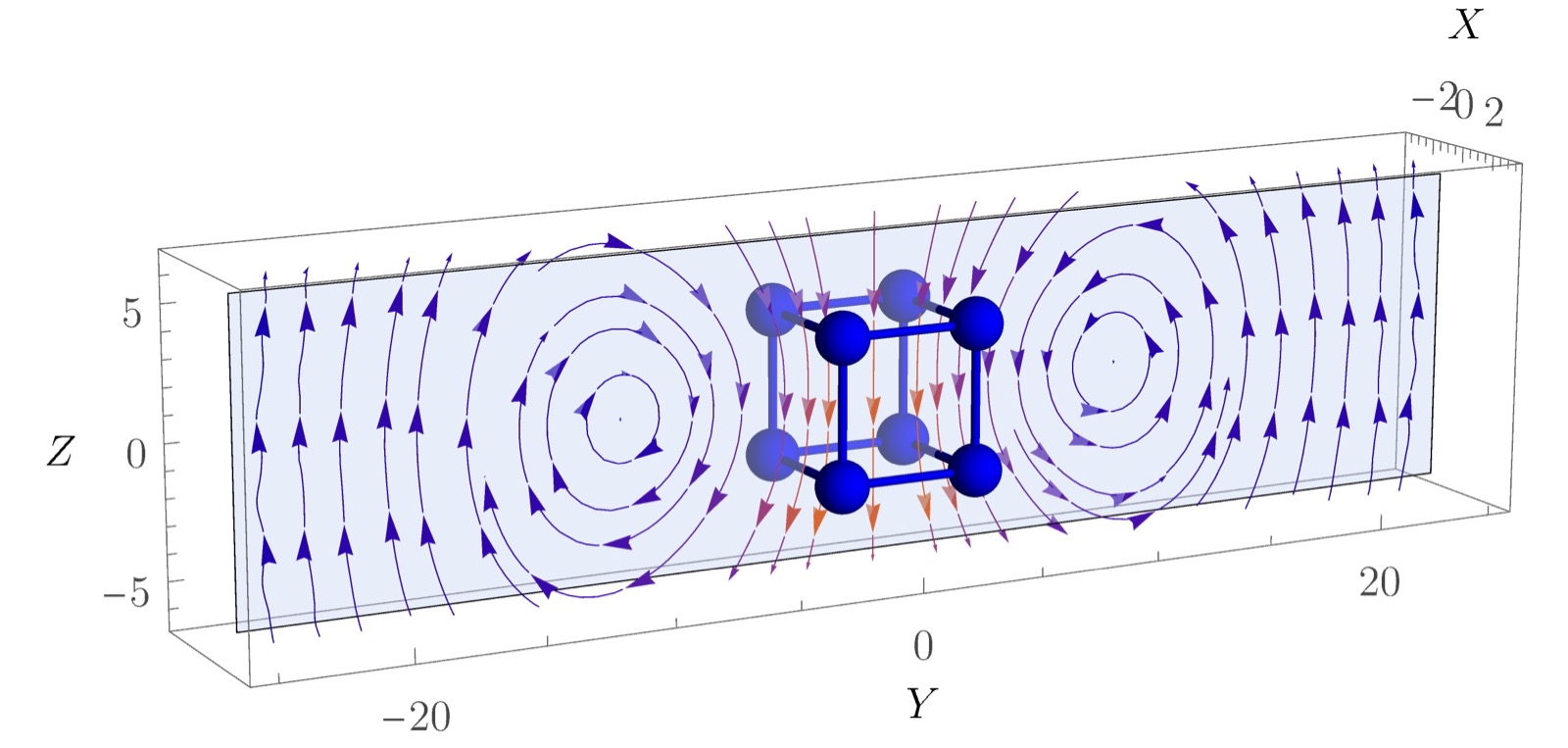}
    \caption{Velocity field lines on the PIV plane (yz plane) near falling SC structure in the bounded container, obtained using Boundary Integral Method (BIM). The finite container leads to backflow, resulting in the recirculating zones, also observed in experiments. The axes are normalised by the radius of the sphere $a$.}
    \label{streamPlot}
\end{figure*}
Here $\oplus$ denotes the direct sum, and $\bm{\mathcal L}^d$ is the double-layer operator given by
\begin{multline}
    \label{eq:Ldeq}
    \bm{\mathcal L}^d(\q)[\x_s] \equiv \bigoplus_{\alpha=1}^N \frac{1}{4\pi}\bigg\{\sum_{\beta=1}^N \oint_{S_\beta}\q_\beta (\x)\cdot \bm{\mathcal T}(\x,\x_s)\cdot \hat{\n}_\beta dS(\x) \\
      -\oint_{C}\q_C (\x)\cdot \bm{\mathcal T}(\x,\x_s)\cdot \hat{\n}_C dS(\x)\bigg\}_{\x_s\in S_\alpha} \\
      \oplus \frac{1}{4\pi}\bigg\{\oint_{C}\q_C (\x)\cdot \bm{\mathcal T}(\x,\x_s)\cdot \hat{\n}_C dS(\x)  \\ -\sum_{\beta=1}^N \oint_{S_\beta}\q_\beta (\x)\cdot \bm{\mathcal T}(\x,\x_s)\cdot \hat{\n}_\beta dS(\x) \bigg\}_{\x_s\in C}
\end{multline}
The rigid body projection operator is given by:
\begin{subequations}
\begin{gather}
    \label{eq:Prbeq}
    \bm{\mathcal P}^{\text{RB}} \equiv \bigoplus_{\alpha=1}^N \bm{\mathcal P}^{\text{RB}}_\alpha \oplus \mathbf{0}, \\
    \bm{\mathcal P}_{\alpha}^{\text{RB}}(\q_\alpha)[\x_s] \equiv \frac{1}{|S_\alpha|} \oint_{S_\alpha} \q_\alpha(\x)dS(\x) \\ \nonumber + \bigg( \mathbf{I}_\alpha^{-1} \cdot \oint_{S_\alpha}\mathbf{X}_\alpha \times \q_\alpha(\x) dS(\x) \bigg) \times \mathbf{X}^s_\alpha, \\
    \mathbf{I}_\alpha \equiv \oint_{S_\alpha} [\bm\delta (\mathbf{X}_\alpha\cdot\mathbf{X}_\alpha)  - \mathbf{X}_\alpha \mathbf{X}_\alpha]dS(\x), \quad \\ \nonumber  \mathbf{X}_\alpha\equiv \x-\x_\alpha, \quad \mathbf{X}^s_\alpha\equiv \x_s-\x_\alpha,
\end{gather}
\end{subequations}
where $|S_\alpha|$ denotes the surface area of the body $\alpha$. The normal projection operator $\bm{\mathcal P}^n$ is given by:
\begin{multline}
    \label{eq:Pneq}
        \bm{\mathcal P}^n(\q \oplus \q_C)[\x_s] \equiv \frac{1}{|S|} \bigg\{ \bigoplus_{\alpha=1}^N \hat{\n}_\alpha(\x_s) \oint_{S_\alpha} \hat{\n}_\alpha\cdot 
        \q_\alpha dS\bigg|_{\x_s\in S_\alpha}  \\
        \oplus \hat{\n}_C(\x_s) \oint_C \hat{\n}_C\cdot \q_C dS\bigg|_{\x_s\in C}\bigg\},
\end{multline}
where $|S| \equiv \sum_{\alpha=1}^N |S_\alpha| + |S_C|$.
The operator on the left-hand side of equation \eqref{eq:qDLeqn} can be inverted to determine $\q \oplus \q_C$ uniquely \cite{KimKarrila2005, Pozrikidis1992}. The rigid body velocities of each object inside the container are then given by
\begin{equation}
    \label{eq:mobQeqn}
    \bm U_\alpha^{\text{RB}} = \bm{\mathcal{P}}^{\text{RB}}_\alpha \q_\alpha, \quad \alpha \in \{1,2,..., N \}.
\end{equation}
This formulation can be used to obtain the mobility matrix for a system of rigid bodies inside a stationary container and the corresponding resistance matrix is simply the inverse of the mobility matrix. Once the double-layer density and mobility of the bodies are obtained from equation \eqref{eq:qDLeqn}, the fluid velocity field can be determined from equation \eqref{velo_BIM}. This velocity field for the SC structure is shown in Fig.~\ref{streamPlot}. Note that the BIM simulations do not take into account the rods connecting the spheres. It imposes the rigid body constraint of the SC structure using the resistance matrix, described in equation \eqref{eq:commonVel}.\\

\section{Results}
\label{results}

\begin{figure*}[t]
    \centering
        \includegraphics[width=\linewidth]{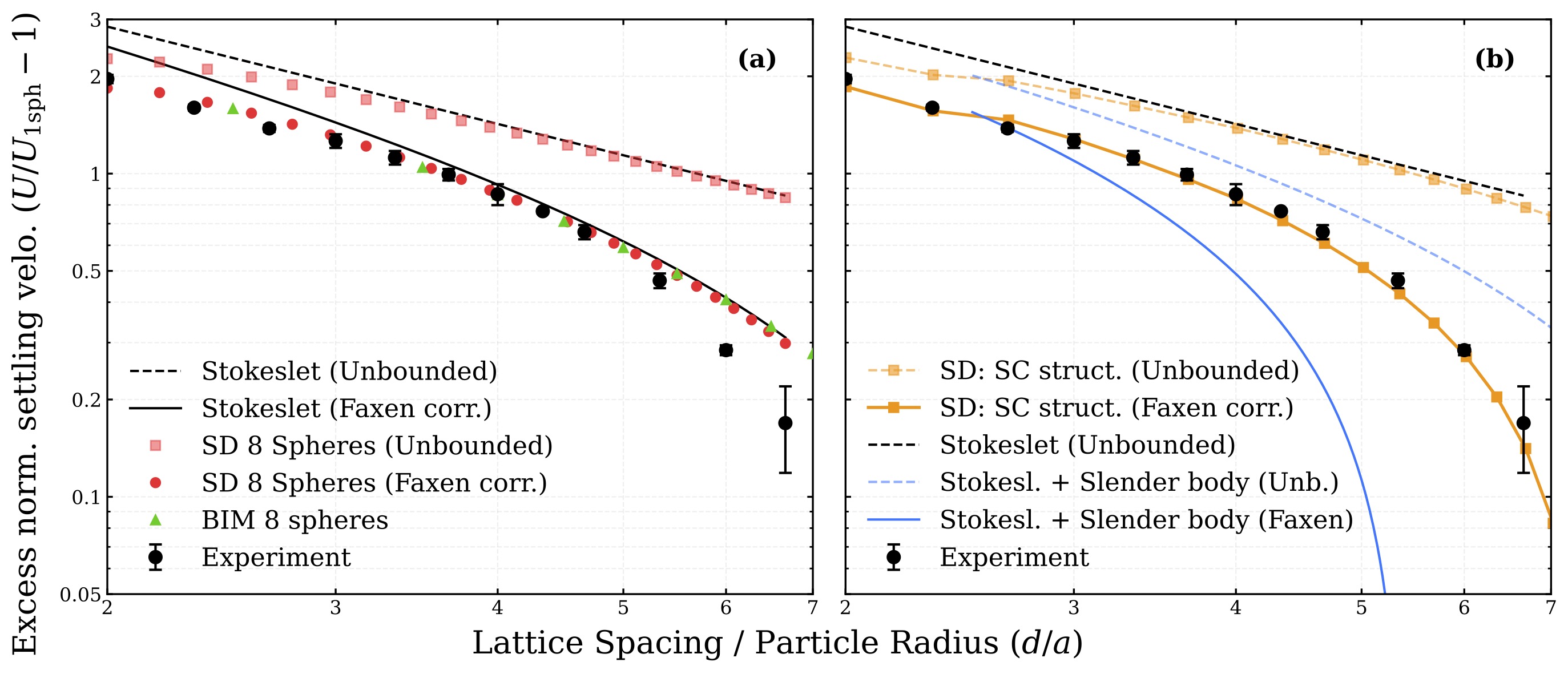}
    \caption{Excess terminal velocity of the SC structure for different lattice spacing $d$, for experimental data and different analytical models. (a) The SD simulation for eight spheres in an unbounded domain show convergence against the Stokeslet prediction \eqref{formula_a_i}. The Boundary Integral Method (BIM) simulations in the bounded domain  and the Fax\'en corrected Stokesian Dynamics simulations for such eight spheres show good agreement between each other, validating the  Fax\'en correction. Furthermore, they align well with experimental data for small $d/a$ ratios. (b) Whereas the Slender-body prediction \eqref{formula_a_ii} deviates substantially from experimental data, SD simulations of the full SC cluster with Fax\'en boundary correction shows excellent agreement with experiments. For SD, the simulation data (squares) are connected by a line to better indicate the good agreement with the experimental data.  }
    \label{fig_sc_all}
\end{figure*}

\subsection{Scaling of terminal velocity, $U$ with solid fraction, $\phi_s$}
The normalized terminal velocity obtained for all seven Bravais lattice shapes is shown in Fig.~\ref{fig_raw_power_law}. We find, remarkably, a power-law dependence of the terminal settling speed, $U$ of all shapes as a function of the solid volume fraction, $\phi_s$, which  is calculated as a fraction of the convex-hull volume occupied by the spheres and the rods (details in appendix~\ref{sec_steiner_formula}). The velocity is scaled by the settling speed that a single sphere would have if it had the same mass as our complex object as discussed through eqn.~\eqref{scaling_exp_2_3}.

We isolate the experimental results for the simple cubic structure and compare it in Fig.~\ref{fig_sc_all}a and \ref{fig_sc_all}b  to different theoretical and numerical results. It is instructive to view these in log-log scale, and to plot the increase in settling velocity as compared to a single sphere. First, we see a significant difference in all cases between the unbounded results and those which are corrected for the existence of the finite container which highlights the effect of the wall on the settling speed. In fact the wall, though far away, is no less effective at changing the dynamics than the immediately neighbouring spheres. This is because of the large-scale recirculating flow that the walls inevitably set up, evident from the BIM calculations in Fig.~\ref{streamPlot}. The Stokeslet approximation is plotted in four curves: (i) with just the spheres in an unbounded domain (eqn. \eqref{formula_a_i}); (ii) with wall corrections (i. and ii. in Fig.~\ref{fig_sc_all}a); (iii) with slender-body accounting of the rods (eqn. \eqref{formula_a_ii}) and (iv) with both corrections (iii. and iv. in Fig.~\ref{fig_sc_all}b). The Stokeslet models do not agree with the experimental results, rendering them not suitable for obtaining reliable settling speeds of complex bodies.

We then examine the SD simulation results with eight spheres and without the slender rods, shown in Fig.~\ref{fig_sc_all}a. First observation is that the settling speed of eight spheres converges to the Stokeslet prediction for high lattice spacings $d$. Second, when we account for the container wall by including Fax\'en corrections, the model agrees well with experiments except at high lattice sizes. Figure \ref{fig_sc_all}a furthermore reveals that the BIM simulations, carried out inside a smoothed approximation of the container with just eight spheres and no rods, have excellent agreement with the Fax\'en-corrected SD. This shows that the  Fax\'en boundary correction \eqref{Faxen_formula} is sufficient to capture the boundary effect of the finite sized container onto the SC structure. Further demonstration of the efficacy of SD is obtained when predictions of the full SC cluster, i.e.  with slender rods, are made by including the Fax\'en corrections. As depicted in Fig.~\ref{fig_sc_all}b, these SD simulations agree well with the experiment at all lattice spacings.

\begin{figure}[t]
    \centering
        \includegraphics[width=\linewidth]{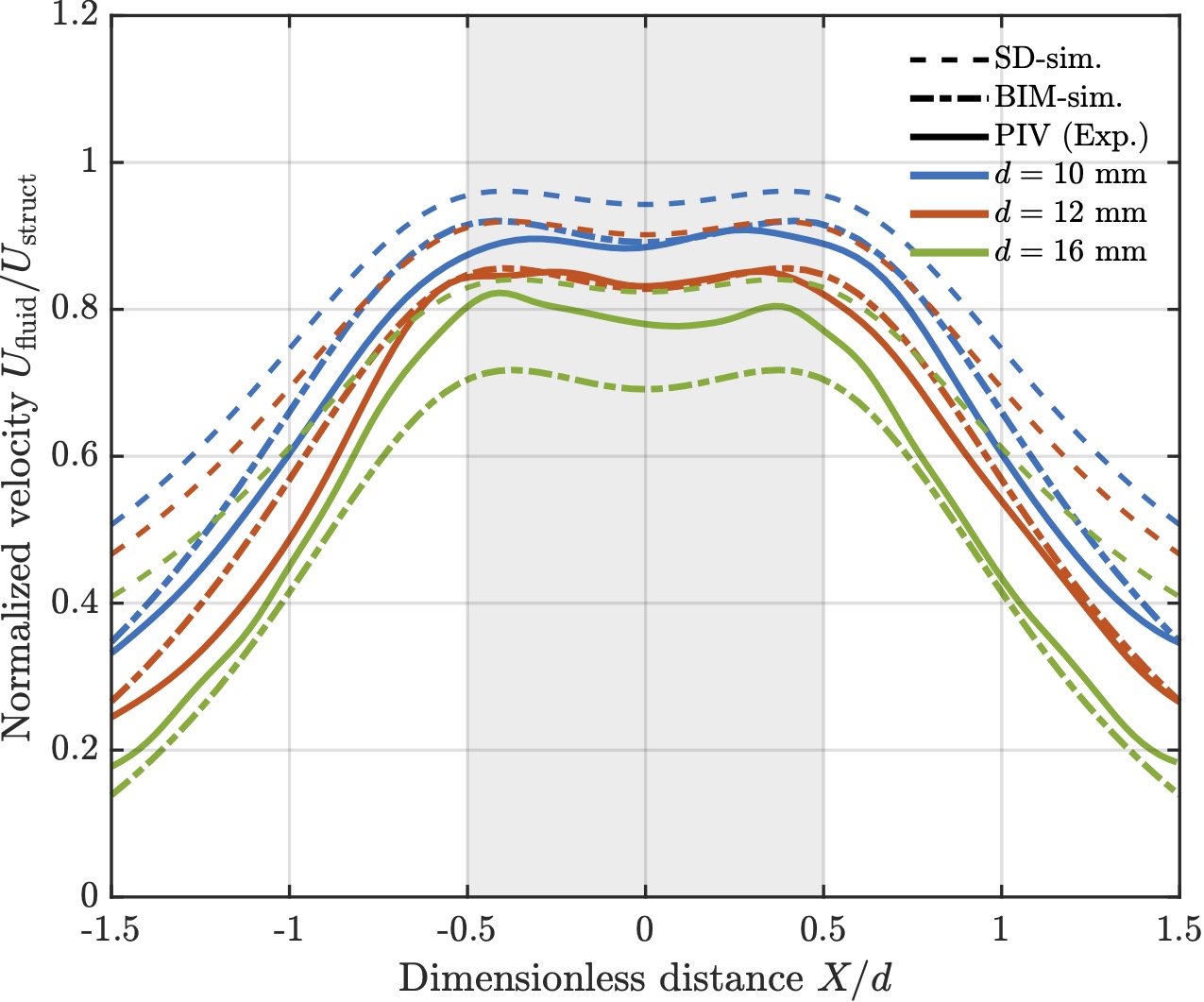}
    \caption{Comparison between velocity profile for a SC structure at the extraction line obtained from PIV experiments (see Fig.\ref{piv_setup}), and SD simulations (with rods, unbounded domain) as well as BIM simulations (without rods, bounded domain). The grey area indicates the interior of the SC structure. The velocity profile is  normalized with the structure's terminal velocity, which depends on $d$.}
    \label{sc_settling_and_piv}
\end{figure}

Having established the applicability of the Fax\'en's boundary correction to the SC structure via the results of Fig. \ref{fig_sc_all}, we return to our main result already depicted in Fig. \ref{fig_porosity_vs_velocity}, the power-law scaling of the normalized velocity as a function of the solid volume fraction: 
\begin{equation}
    U\propto \phi_s^{0.30} \cdot  U_\mathrm{1sph} \cdot \left(\frac{m_\mathrm{struct}}{m_\mathrm{1sph}}\right)^{2/3}.
    \label{eqn_phi_scaling}
\end{equation}
The experimentally less intensively tested structures FCC, octahedron and HCP (see Table \ref{Tab_Strcuture_Data} for the respective numbers of different lattice constants investigated)  show a certain deviation from the exponent $0.30$, whereas all the well-tested lattice types, namely tetrahedron, BCC and SC, agree in their prediction of $0.30$ for the scaling exponent.\\

If the outer shape were a perfect sphere, then both in the low-porosity limit and the high-porosity limit, we obtain a power-law for the settling velocity, with an exponent of $1/3$ \cite[eq. C11]{Lasso1986}. This is known from the Darcy-Brinkman model \cite{brinkman1947, HappelBrenner1983} of settling spherical porous objects. However, the same power-law with matching exponents for differently shaped bodies has not, to our knowledge, been observed before.

\subsection{Flow field inside the container}
\label{sec_piv_results}

\begin{figure}[t]
 \centering
 \includegraphics[width=\linewidth]{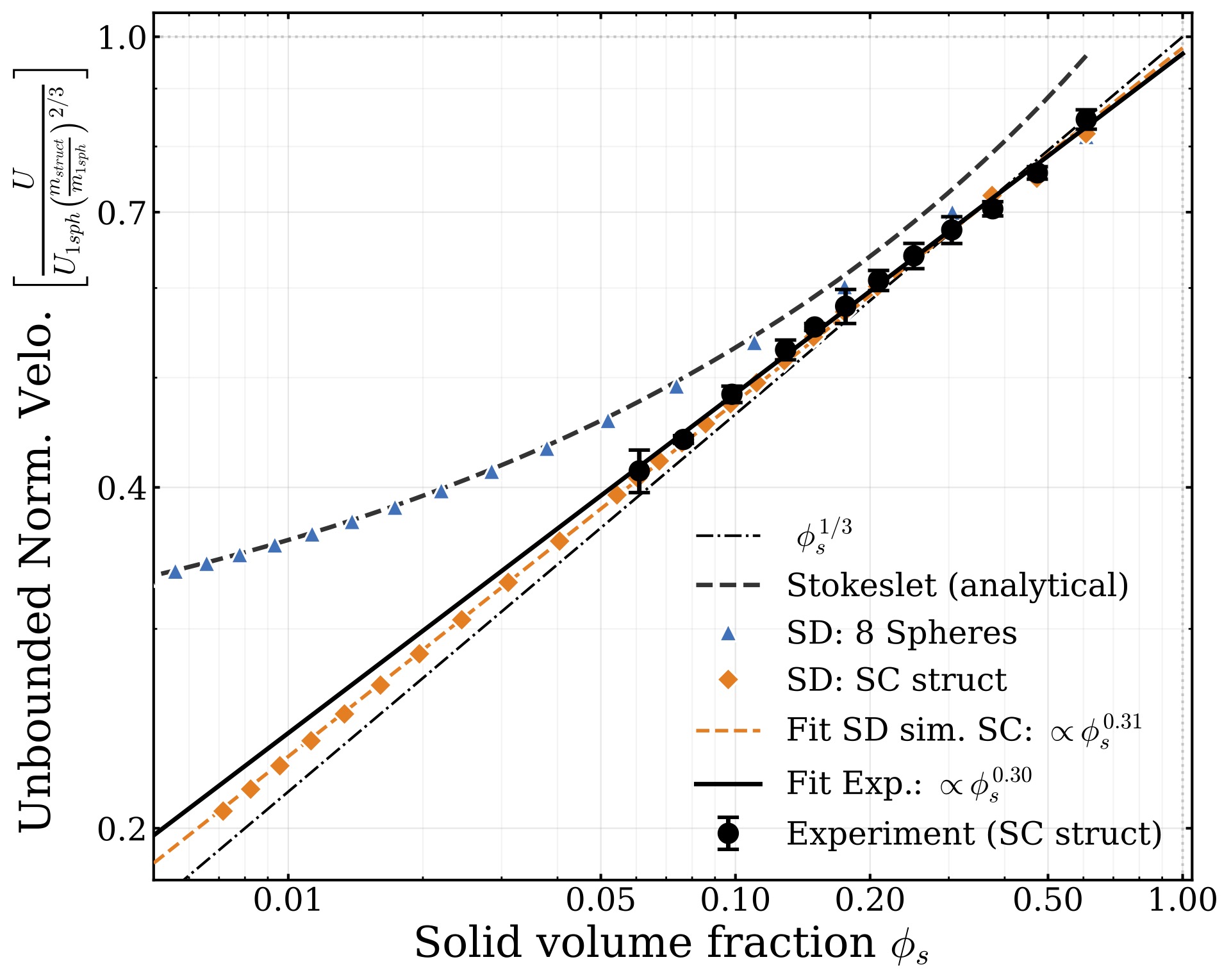}
 \caption{Influence  of connecting rods on the normalized settling speed as a function of $\phi_s$: Stokesian Dynamics simulation of eight spheres with and without rods vs experimental values. Note that the SD simulations suggest that the power-law behaviour continues to hold for the whole  SC structure, the eight spheres without rods align well with the Stokeslet for small $\phi_s$, with $U_\mathrm{8spheres}\rightarrow{} U_\mathrm{1sph}$ as $\phi_s \rightarrow 0$, hence converging to $0.25$ on the $y$-axis.}
 \label{sc_struct_vs_8_spheres}
\end{figure}

The internal flow field of the SC structure at different lattice spacing $d$ were obtained in experiments along a particular extraction line (shown in Fig.~\ref{piv_setup}(a)) and is shown in Fig.~\ref{sc_settling_and_piv}. The measured flow field is another measurable that highlights the importance of considering the boundary effects due to the container. We find that BIM simulations, for $d= 10\ \mathrm{mm}$ and $d= 12\ \mathrm{mm}$, are in good agreement with experimentally obtained data, whereas SD simulations consistently predict higher velocities than measured in the experiments. Note that the velocities depicted in Fig.~\ref{sc_settling_and_piv} are relative to the structure's velocity. The discrepancy of the BIM simulations at $d=16\ \mathrm{mm}$ is likely an effect of the missing rods in BIM simulations, as the rods tend to confine the fluid inside the SC structure and hence  increase the relative speed of the fluid.

In summary, our experiments and numerics reveal the relationship between settling velocity and the porosity of structured objects such as Bravais lattice unit-cells and the importance of the boundary effects while calculating settling velocity. Our BIM and SD simulations with the Fax\'en corrections that account for the wall-effects are sufficient to capture the velocity far away from the object. Further, our detailed comparisons with experiments, BIM and SD simulations indicate that arbitrary shaped objects may be adequately represented by an appropriate arrangement of spheres of different radii, and SD performed on this collection of spheres provides good prediction for the settling velocity. We hope that our results on the settling of lattice structures provides impetus for further studies on more complex geometries such as ice-crystals in clouds, micro-plastics, phytoplanktons and biologically generated detritus sinking through the ocean.

\section{Discussion}

Our study shows that the settling speed of rigid lattice structures exhibits a universal power-law dependence on the solid fraction $\phi_s$, with  internal microstructural details contributing only to the prefactors. However, our detailed analysis of different models for Stokesian settling indicates that the functional dependence of the settling velocity on $\phi_s$ is sensitive to the fine-scale features and boundary effects. 
Specifically, we find that the presence of thin connecting rods  in the unit-cells substantially modifies the settling velocity in an unbounded domain. Stokesian Dynamics simulations predict an extended regime over which the experimentally observed scaling law remains valid, notably extending its applicability to smaller values of $\phi_s$ than experimentally measured, as shown in Fig. \ref{sc_struct_vs_8_spheres}.
Such high-porosity regimes are particularly relevant in natural contexts, for instance, in aggregates of phytoplankton in marine environments \cite{Flintrop2023, Kirboe2001}. It is therefore plausible that, while loosely packed, weakly connected  phytoplankton aggregates with negligibly low mucus content may settle in accordance with the Stokeslet prediction, whereas rigid interconnections between constituent units, such as those arising from diatom chains [see Fig 2B in \cite{Kirboe2001}], can significantly reduce their settling velocity. Note that our power-law scaling does not capture the presence of polymeric fluid or gel phases in such aggregates \cite{Chajwa2024}.

Furthermore, we observe a small but systematic deviation of the SD predictions from the $\phi_s^{1/3}$ scaling at lower values of $\phi_s$, even though the agreement between the SD simulations and the $1/3$ scaling law remains remarkably strong, as visible in Fig. \ref{fig_porosity_vs_velocity}. This is especially striking given that the derivation of the $1/3$ exponent in \cite[Eq. C11]{Lasso1986} assumes an impermeable structure $-$ a condition clearly violated in the high-porosity limit considered here.
A key advantage of the present experimental framework lies in the precise tunability and computability of the solid volume fraction, or equivalently, the porosity. In addition, the use of fixed rigid-body configurations ensures a high degree of reproducibility, in contrast to the random agglomerations considered in earlier studies (e.g. \cite{Alabrudzinskietal2009}).

The experiments presented here are limited to regular geometries, and it remains to be tested whether the observed scaling of settling speed extends to more irregular arrangements of spheres connected by rods. Such structures can be analysed within the present framework, as the volume of the convex hull may be computed from the sphere positions and radii via Steiner’s formula. A key difference is that irregular geometries may induce rotational dynamics during settling, in contrast to the symmetric structures studied here, necessitating consideration of orientation-dependent settling speeds. This raises two related questions: whether a power-law scaling persists for irregular structures, and, if so, whether the corresponding exponent remains consistent with the value $\gamma = 0.3$ obtained here.

\appendix

\section{Computations of solid volume fraction}
\label{sec_steiner_formula}

\begin{table}
\centering
\begin{tabular}{@{}lccc@{}}
\toprule
\textbf{Structure} & \textbf{$V_{\mathrm{poly}}$} & \textbf{$S_{\mathrm{poly}}$} & \textbf{$\sum l_i \alpha_i$} \\ \midrule
SC                 &  $d^3$ & $6d^2$ & $6\pi d$ \\
BCC                &  $d^3$ & $6d^2$ & $6\pi d$ \\
FCC                 & $d^3$ & $6d^2$ & $6\pi d$ \\
Tetrahedron        & $\frac{\sqrt{2}}{12}d^3$ & $\sqrt{3}d^2$ & $11.46 \ d$ \\
Sq. Pyramid  & $\frac{1}{6}d^3$ & $(1+\sqrt{2})d^2$ & $\pi\left(3 + \frac{4}{\sqrt{3}}\right)d$ \\
Octahedron        & $\frac{\sqrt{2}}{3}d^3$ & $2\sqrt{3}d^2$ & $14.77 \  d$ \\
HCP     & $\frac{\sqrt{2}}{2}d^3$ & $\left(\frac{\sqrt{3}}{2} + 2\sqrt{6}\right)d^2$ & $\pi\left(3 + 4\sqrt{\frac{2}{3}}\right)d$ \\ \bottomrule
\end{tabular}
\caption{Terms in Steiner's formula in eqn.~\eqref{eqn_Steiner} used to compute the volume of the Convex Hull enclosing the structure. $N_s$ is the number of spheres, $N_c$ the total number of rods connecting the spheres.}
\label{Tab_Steiner_coeffs}
\end{table}

In Fig.~\ref{fig_porosity_vs_velocity}, the solid volume fraction $\phi_s$ refers to the ratio of volume occupied by spheres and rods divided by the total convex hull's volume of the structure, i.e. $\phi_s= V_\mathrm{solid}/V_{\mathrm{ConvHull}} $.  In order to determine this value, Steiner's formula
\begin{equation}
    V_{\mathrm{ConvHull}} = V_{\mathrm{poly}} + S_{\mathrm{poly}} a + \frac{1}{2}\left(\sum l_i \alpha_i\right) a^2 + \frac{4}{3}\pi a^3,
    \label{eqn_Steiner}
\end{equation}
is applied to every structure. Here, $a$ is again the sphere's radius and $V_{poly}$ is the volume of a structure when considering the convex hull obtained from all centres of the spheres alone, making it a polyhedron where only surfaces and edges at the boundary are retained (important in the HCP case). $S_{\mathrm{poly}}$ is the sum of the areas of the surfaces of this polyhedron, and $l_i$ is the set of its edges, accompanied with the exterior angle $0\leq\alpha_i \leq \pi$ that the planes connected by $l_i$ span. Table \ref{Tab_Steiner_coeffs} gives the relevant terms within Steiner's formula for the seven structures under investigation.

\medskip
\hskip -5mm {\bf AI Assistance Statement} 

Code assistance by Gemini was partially used for the following aspects of the work: 1. Improvement of figure layout/design (Fig. 1a, 2a, 3, 4, 5, 7, 8, 9) . 2. Stokesian Dynamics simulations and  PIV data analysis (code-generation and debugging, manually checked and validated when used).


\medskip

\hskip -5mm {\bf Acknowledgements} We acknowledge support from the Department of Atomic
Energy, Government of India, under Project No. RTI4001. SB would like to thank the DAAD Rise program through which the research visit at the ICTS in Bengaluru was made possible.  RG thanks the ANRF JC Bose Grant for support during the recent part of this work.

\medskip

\hskip -5mm {\bf Dedication} We dedicate this paper to the memory of Professor KR Rajagopal. We have tried to do justice to his penchant for correctness, by performing several different levels of theoretical and numerical study, and carrying out careful experiments. RG wishes to acknowledge years of discussions with KRR, and important encouragement during her early career.

\vskip2pc

\bibliographystyle{apsrev4-1}
\bibliography{references}
\end{document}